\documentclass[3p,times,twocolumn]{elsarticle}

\usepackage{ecrc}
\usepackage{amssymb,amsmath,epsfig}

\volume{00}

\firstpage{1}

\journalname{Nuclear Physics B Proceedings Supplement}

\runauth{Albrecht Karle}

\jid{nuphbp}

\jnltitlelogo{Nuclear Physics B Proceedings Supplement}
\usepackage{amssymb}
\usepackage[figuresright]{rotating}

\begin{document}

\begin{frontmatter}
\dochead{}

\title{Neutrino Astronomy - A Review of Future Experiments }

\author{Albrecht Karle}

\address{Department of Physics and Wisconsin IceCube Particle Astrophysics Center, University of Wisconsin-Madison, Madison, WI 53706, USA}

\begin{abstract}
Current generation neutrino telescopes cover an energy range from about 10 GeV to beyond $10^9$ GeV. 
IceCube sets the scale for future experiments to make improvements. 
Strategies for future upgrades will be discussed in three energy ranges.  At the low-energy end, 
an infill detector to IceCube's DeepCore would add sensitivity in the energy range 
from a few to a few tens of GeV with the primary goal of measuring the neutrino mass hierarchy. 
In the central energy range of classical optical neutrino telescopes, next generation
detectors are being pursued in the Mediterranean and at Lake Baikal.  The KM3NeT 
detector in its full scale would establish a substantial increase in sensitivity over IceCube. 
At the highest energies, radio detectors in ice are among the most promising and 
pursued technologies to increase exposure at $10^9$ GeV by more than an order of 
magnitude compared to IceCube.

\end{abstract}

\end{frontmatter}

\section{Introduction}
\label{sec:introduction}

Highest energy cosmic rays provide evidence of the existence of powerful 
accelerators in the Universe.  
Cosmic particles have been observed up to energies beyond $10^{20}$ eV. 
One hundred years after the first discovery of cosmic rays, their origin 
remains largely unresolved. 
Point sources of high-energy gamma rays have been observed up to energies 
of about 100 TeV.  They provide evidence of accelerators of energetic radiation. 
Yet, the connection to the cosmic rays of higher energies is unclear
and the high-energy photon view to the Universe is blocked due to interactions 
with low-energy photons, the microwave background and extragalactic 
photon backgrounds. 
Neutrinos may traverse the Universe even at their highest energies.
Cosmic ray interactions, either in the accelerator regions or at higher energies
on the same photon background, will inevitably produce neutrinos at some level.
Thus, neutrino astronomy may provide the missing clues to uncovering the origin of cosmic rays.

The IceCube neutrino observatory, the largest detector with 5160 optical sensors
deployed in a billion tons of ice,  has been in full operation since May 2011 and has already accumulated an 
unprecedented exposure to cosmic neutrino sources and atmospheric neutrinos 
from 10 GeV to $10^9$ GeV.  
Neutrino telescopes, while initially not designed to probe lower energy
atmospheric neutrino oscillations, have started to explore the energy 
range from 10 to 100 GeV.  IceCube \cite{andreas_poster,Sullivan_Nu2012} and 
the ANTARES \cite{Antares_oscillations} experiment are reporting first observations of atmospheric 
neutrino oscillations with their detectors.  

IceCube is now setting the benchmark for future detectors. 
It has already accumulated more exposure to high-energy neutrinos, from 10 GeV to 
beyond 1 EeV, than any other experiment.  
A very detailed review of the current state of neutrino astronomy can be found in \cite{SpieringKatz}.  
We will discuss three energy scales with respect to expansion beyond the current detector capabilities. 

\subsection {Low-energy extensions: 1 to 100 GeV}
At low energies, IceCube has already improved its sensitivity compared to 
the baseline envisioned at its outset \cite{icecube_performance_paper2002}
by adding a central dense infill named DeepCore \cite{DeepCorePaper}. 
The primary strategy was to arrange additional strings optimized for lower energies 
around 10 GeV in the bottom center of IceCube. 
The DeepCore infill detector relies on the main IceCube array to function 
as a veto against cosmic-ray muons which trigger IceCube
at a rate of about 3 kHz at a depth of 2 km. 
Preliminary results on neutrino oscillation measurements have 
been presented and the increased sensitivity to dark matter has been presented 
as well. Building on this, studies are now underway to add an additional infill 
array to lower the threshold to a few GeV. 
The primary science goal that such a dense detector 
is envisioned to address is the mass hierarchy problem of neutrinos.

\begin{figure}
\begin{center}
	\resizebox{\linewidth}{!}{\includegraphics{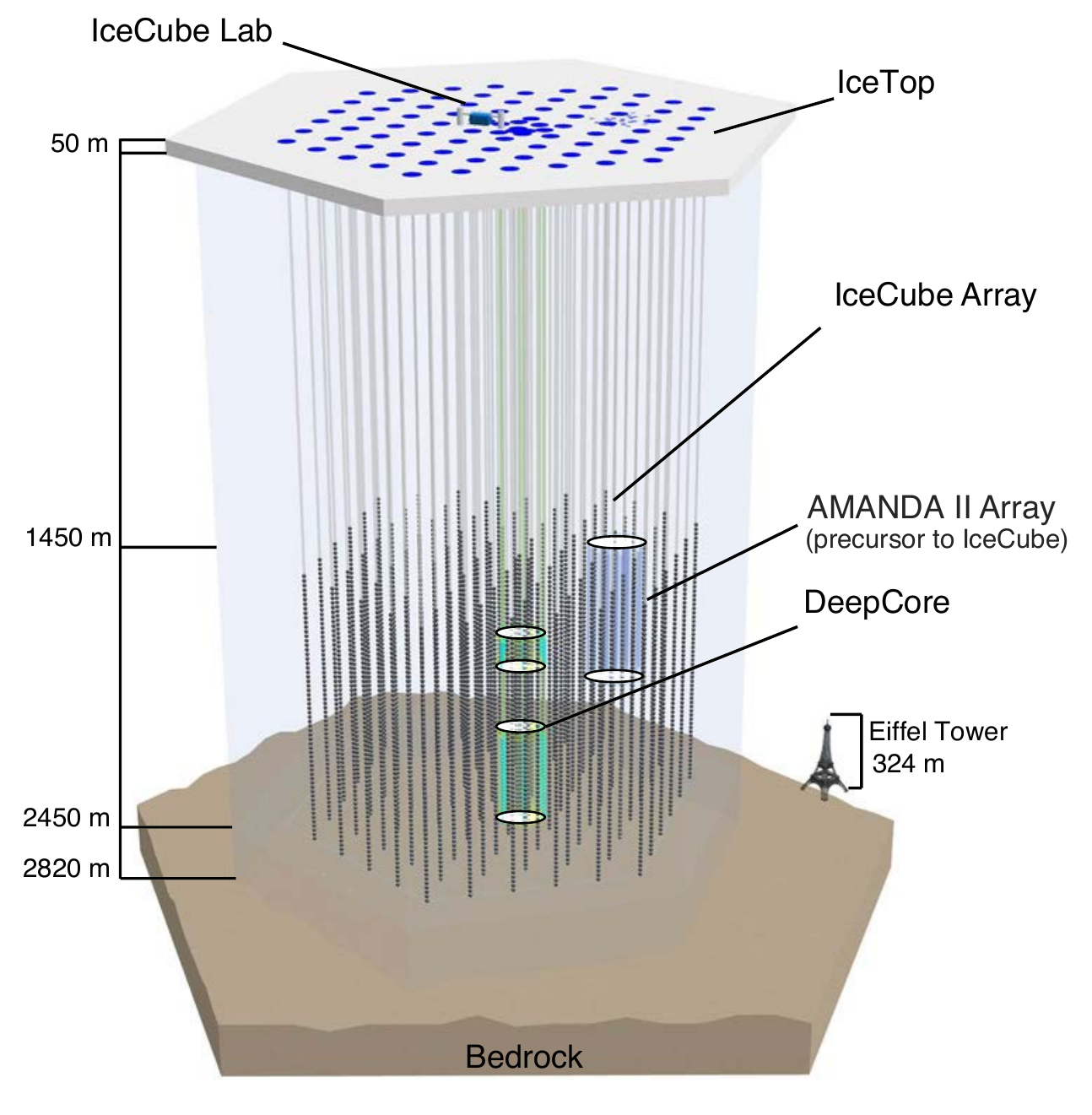}}
	\caption{Schematic view of IceCube. 5160 sensors are deployed on 86 strings between 
	1450 and 2450\,m depth. An additional 324 sensors are deployed in the IceTop surface 		detector.  }
	\label{fig:IceCube}
\end{center}
\end{figure}

\subsection {Neutrino astronomy from TeV to beyond 10 PeV energies}
The energy scale from 1\,TeV to 10\,PeV may be considered the 
classical energy scale for neutrino astronomy. 
The lower end of this energy range allows searching 
for steeper energy spectra such as from galactic sources. 
Harder energy spectra, such as a neutrino flux 
emerging from cosmic-ray acceleration sites (e.g., the Waxman-Bahcall 
flux \cite{WaxmanBahcall} with a harder spectral index of E$^{-2}$), 
require detectors to be effective at energies well beyond 100\,TeV
where the atmospheric neutrino background fades. 
The KM3NeT consortium \cite{km3net} is planning to 
build such a telescope in the Mediterranean Sea. 
It would exceed IceCube both in deployed photon detection 
area as well as in neutrino effective area by a factor greater than 5. 
It would vastly increase the sensitivity to neutrino sources in the galactic center region, 
especially at TeV energies.

\subsection {Cosmogenic highest energy neutrinos: 100 PeV to 100 EeV}
Cosmic rays have been measured to energies beyond $10^{20}$ eV.
At these extreme energies cosmic-ray protons interact with the 
cosmic microwave background photons to produce pions and 
a predictable neutrino flux in the primary energy range from  $10^{17}$ to 
 $10^{19}$ eV. 
The energies are high enough that the radio technique of 
measuring the coherent radio emission from the showers 
produced in these interactions promises a more cost effective 
way to detect neutrinos than the optical Cherenkov-light detection method.

\begin{figure}
\begin{center}
	\resizebox{\linewidth}{!}{\includegraphics{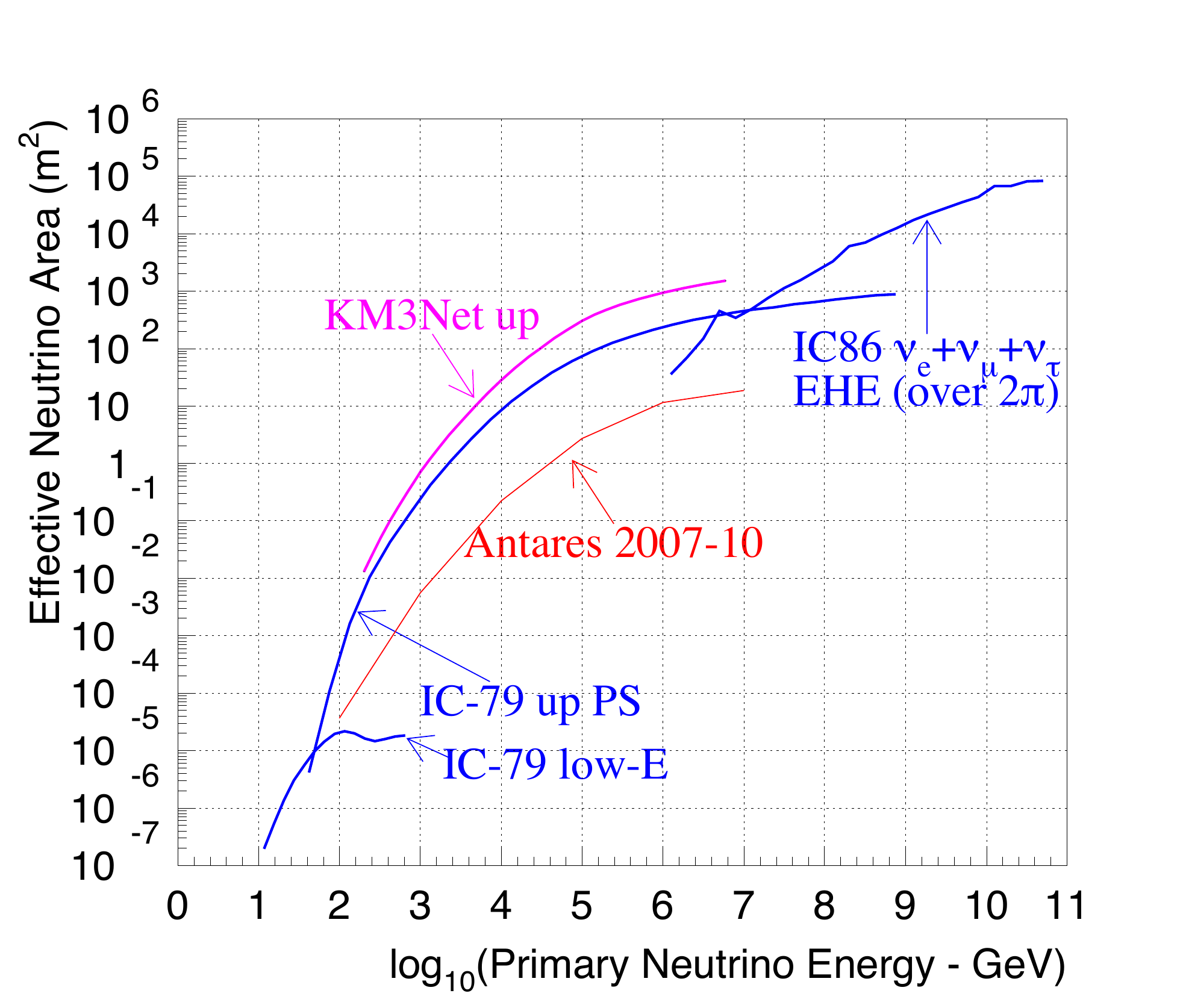}}
	\caption{Effective areas of selected neutrino telescopes.  The strong increase of the effective detector area allows neutrino telescopes to cover a wide range of energies. 
	Effective areas for different event selections are given for IceCube and for ANTARES \cite{AntaresPointsource2012}. Also shown are projections for a design of the KM3NeT detector \cite{km3net}.   }
	\label{fig:effarea}
\end{center}
\end{figure}

Figure \ref{fig:effarea} shows muon neutrino effective areas of selected neutrino telescopes.   
Neutrino effective areas correspond to the area at which the detector would be 
100\% sensitive to an incoming neutrino flux of a given energy. 
The effective area rises strongly with energy due to the growing neutrino-nucleon 
cross section as well as the increase in the range of muons. 
This allows a wide energy range for such telescopes even with uniform geometry. 
At low energies, one can see the effect of the denser instrumentation of
IceCube's DeepCore.  An analysis associated with this effective area was presented 
at this conference \cite{andreas_poster,Sullivan_Nu2012}. 
At energies above 1 PeV, where Earth absorption dominates,
 a different event selection allows the use of downgoing events. 
An effective area for KM3NeT based on current design studies \cite{km3net} is also shown. 
Effective areas of existing detectors are all shown for analysis-level event selections. 
The effective area for KM3NeT is given for an event selection that is considered 
suitable for analysis. 

Future extensions are envisioned for the three energy scales, as discussed in 
more detail in the following sections: 
\begin{itemize}
\item 1 - 100 GeV: Upgrade of IceCube/DeepCore with more dense strings in the center.
\item 100 GeV to 100 PeV:  planned water/ice-based neutrino telescopes, KM3NeT and others.  
\item 100 PeV to 100 EeV: Radio-based neutrino detectors using Antarctic ice.
\end{itemize} 



\section{Future water/ice Cherenkov-based neutrino telescopes}
\label{sec:km3net}

The largest neutrino telescope, IceCube, came into full operation 
in 2011, with 5160 PMTs instrumented in the deep glacial ice at the South Pole.   
IceCube has  already accumulated more than 3\,km$^{2}\cdotp$years of exposure. 
In the next few years it will push the search for cosmic neutrinos, 
the search for dark matter, as well as the measurements made with atmospheric neutrinos 
and cosmic rays to unprecedented exposure and precision. 
In the Northern Hemisphere the ANTARES experiment has been taking data 
and has demonstrated the ability to reach excellent angular resolution and 
a high sensitivity with a water-based instrument of 900 PMTs. 
The Lake Baikal detector in particular has demonstrated the ability to detect 
noncontained cascade events with a large volume and as a result was able 
to obtain diffuse limits on astrophysical neutrinos \cite{Baikal_Diffuse}.

\begin{figure}
\begin{center}
	\resizebox{0.9\linewidth}{!}{\includegraphics{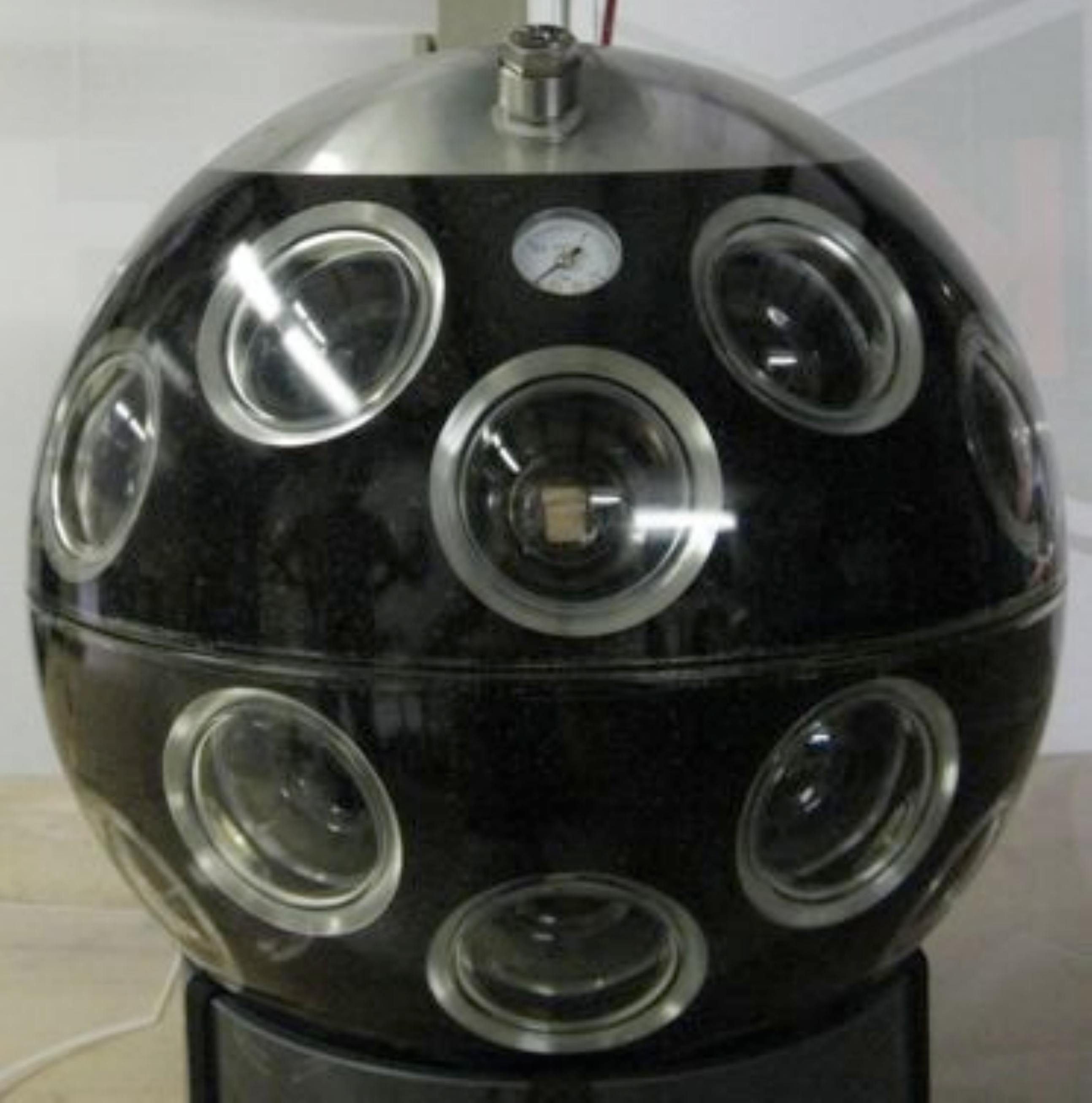}}
	\caption{The optical sensor for KM3NeT contains 31 small PMTs 
	(Photo courtesy KM3NeT\cite{km3net}) }
	\label{fig:KM3NetModule}
\end{center}
\end{figure}

The most developed and ambitious plan for a future experiment is being pursued 
by the KM3NeT consortium.  KM3NeT \cite{km3net} is a proposed neutrino detector 
that builds on the heritage of ANTARES (see results presented at this 
conference \cite{Coyle_Nu2012} and other experimental efforts aimed at constructing
 a very large neutrino detector in the Mediterranean Sea.  
In its current design concept the detector would consist
of 12,800 optical modules on 610 strings covering an instrumented volume 
of approximately 5\,km$^3$. 
The optical modules are based on a novel approach of integrating  
a sizable number of 31 small PMTs of about 75 mm cathode diameter
into one optical sensor. 
Eventually these PMTs are assumed to employ high quantum efficiency 
of about 35 percent. 
A picture of a prototype multi-PMT module is shown in figure \ref{fig:KM3NetModule}. 
Advantages of pixelization include directional sensitivity in all directions, 
signal-to-noise improvements in a relatively high light background of
ocean water, which is of order 10$^3$ times higher than for modules in ice, 
and more opportunity for single photon counting.
In order to get a full sense of the scale of instrumentation 
of this experiment it is necessary to take into account
that the total photon detection area of such a module is significantly larger 
than that of an IceCube Digital Optical Module (DOM).
When considering the PMT coverage of water-Cherenkov detectors
it is useful to define an effective photon detection area, $A_{EPD}$, which takes into account 
quantum efficiency (QE) and collection efficiency (CE) of PMTs:

\begin{center}
$A_{EPD} = A_{cathode} \times QE \times CE $
\end{center}

By this measure the total $A_{EPD}$ of a KM3NeT module is more than 
3 times larger than that of an IceCube DOM and the
total detector effective photo coverage is about 8 times IceCube's.
It is worth mentioning that the KM3NeT plans to transmit all PMT signals
to the shore at a data rate of approximately 1 TB/s before being reduced by 
a triggering and filtering CPU farm to about 10 Mb/s. 

The primary scientific goal of KM3NeT is the detection and measurement of 
neutrino fluxes from galactic sources.  It would allow independent observations
of possible IceCube discoveries with improved significance 
within a reasonable amount of time, and also offers a broad program of science
topics associated with oceanography, geophysics, and marine biology.
 
The Northern Hemisphere location
provides an optimal view of the Galactic Center region, where many 
TeV gamma-ray sources have been detected by ground based air Cherenkov telescopes. 
The optical properties of the deep seawater at the locations considered have been studied 
in detail and have been incorporated into the detector design studies. 
Design studies predict a very good angular resolution a little above $0.1^\circ$ for 
an assumed energy spectrum of $E^{-2}$.  
String spacing and geometry are not final yet.  The instruments are being considered
for deployment at more than one site location. 
Plans are in place for a phase 1 installation. 
The first multi-PMT optical module is planned for installation this summer at the 
ANTARES site and the first phase of construction is scheduled to start later this 
year in Italy and France, with completion envisioned by 2020.

The Baikal Collaboration has been operating the neutrino detector configuration NT200 
in Lake Baikal since 1998. 
Baikal has been using the frozen lake in the winter as an installation and maintenance platform. 
The NT200 configuration spans 72\,m in height and 43\,m in diameter and has been successful 
in reconstructing non-contained events to obtain a sizeable volume for 
the detection of cascade events. 
Baikal is pursuing an upgrade to a "Gigaton Volume Detector" (GVD) \cite{Baikal_GVD1,Baikal_GVD2}. 
It will consist of strings that are grouped in clusters of eight with 
24 PMTs on each string.  The PMTs are foreseen to be of 25\,cm diameter with about 35\%
quantum efficiency.
The GVD configuration expects a total of 2304 optical modules on 96 strings.  
The collaboration envisions an eventual configuration (GVD-4) with 10,368 PMTs on four GVD clusters. 
Based on design studies, the effective volume for cascade events is 0.4 (0.6)\,km$^3$
above 10 TeV (1PeV) for GVD and 4 times as much for GVD-4. 
The muon effective areas range from 0.3 to 1.8\,km$^2$ for the 
considered configurations and energy ranges.

\section{Low-energy extensions}
\label{sec:low_energy}

\begin{figure}
\begin{center}
	\resizebox{\linewidth}{!}{\includegraphics{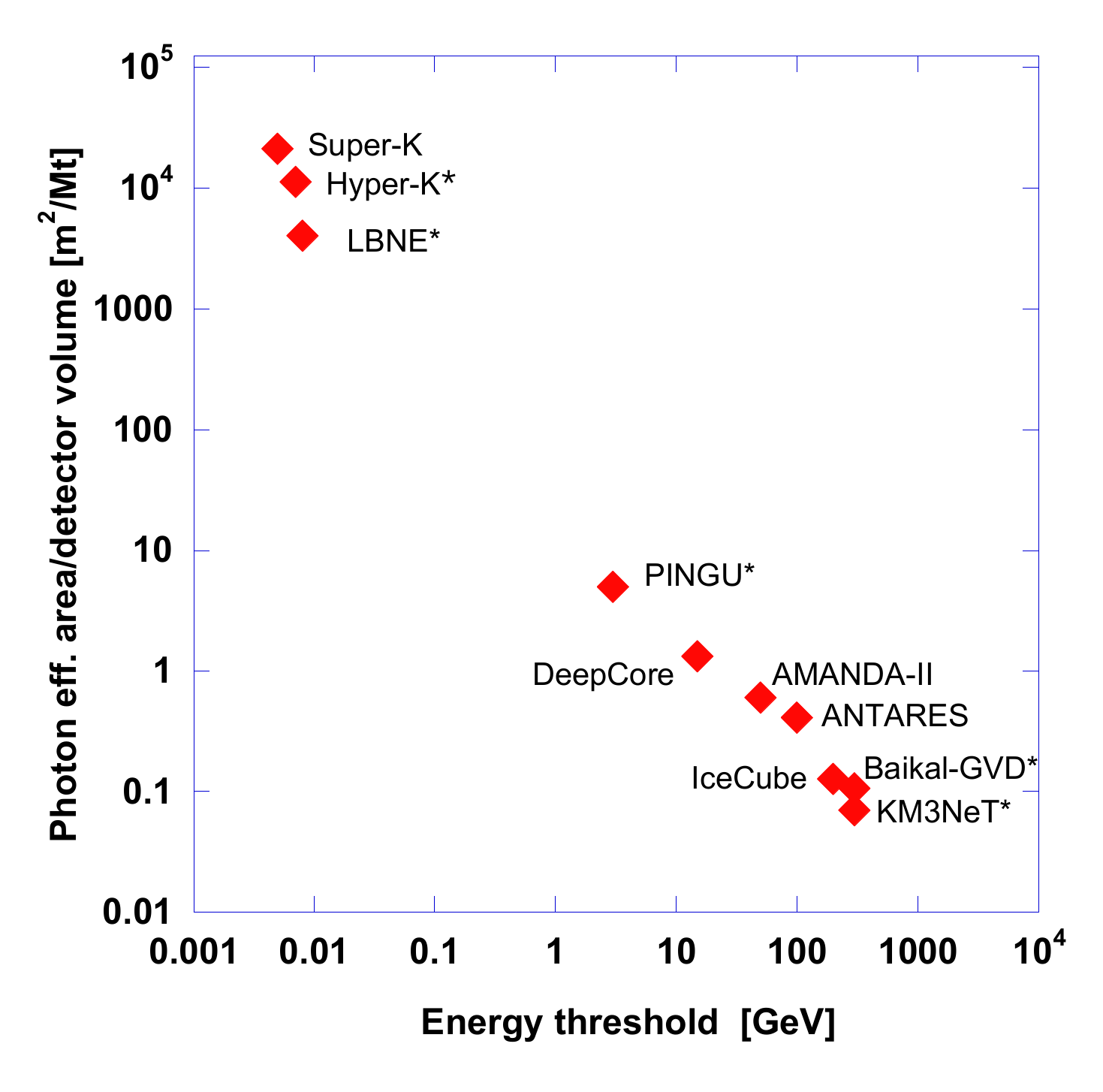}}
	\caption{The effective photodetection area per unit mass is plotted versus 
	the approximate energy threshold of various water/ice Cherenkov detectors.  
	Detectors that are in a planning or a conceptual phase are indicated with an (*). The effective PMT coverage 
	scales roughly to the inverse of the threshold for a wide range of detectors. 
	There are no existing detectors with intermediate thresholds in the range 0.1 to 10 GeV. }
	\label{fig:PMTcoverage}
\end{center}
\end{figure}

As implied by the name neutrino telescope, the primary energy range of very large water/ice Cherenkov detectors has been motivated by astrophysical goals
associated with TeV energies.
The search for dark matter as well as the ability to explore atmospheric neutrino oscillation 
physics with large effective volumes has increased the interest to push the energy threshold of 
"TeV" neutrino telescopes lower. 
The ANTARES collaboration presented an analysis showing successful
observation of neutrino oscillations \cite{Antares_oscillations} 
IceCube deployed a configuration of denser strings, DeepCore, in the central lower
part of the detector for increased sensitivity at low energies. 
Figure \ref{fig:PMTcoverage} shows the increased neutrino effective area at the low energies in 
an initial oscillation analysis of this data stream \cite{andreas_poster}.  

It is interesting to note a significant gap in energy coverage of existing 
neutrino experiments.  
Underground neutrino detectors are typically designed for an energy threshold of 
around 5 to 10\,MeV.  While this is characteristic of many neutrino detectors, 
here we will focus on the water/ice Cherenkov detectors only.
Neutrino telescopes are designed to be
fully efficient for TeV muons and higher energy cascades, a jump of 5 orders of magnitude. 

\begin{figure}
\begin{center}
	\resizebox{0.9\linewidth}{!}{\includegraphics{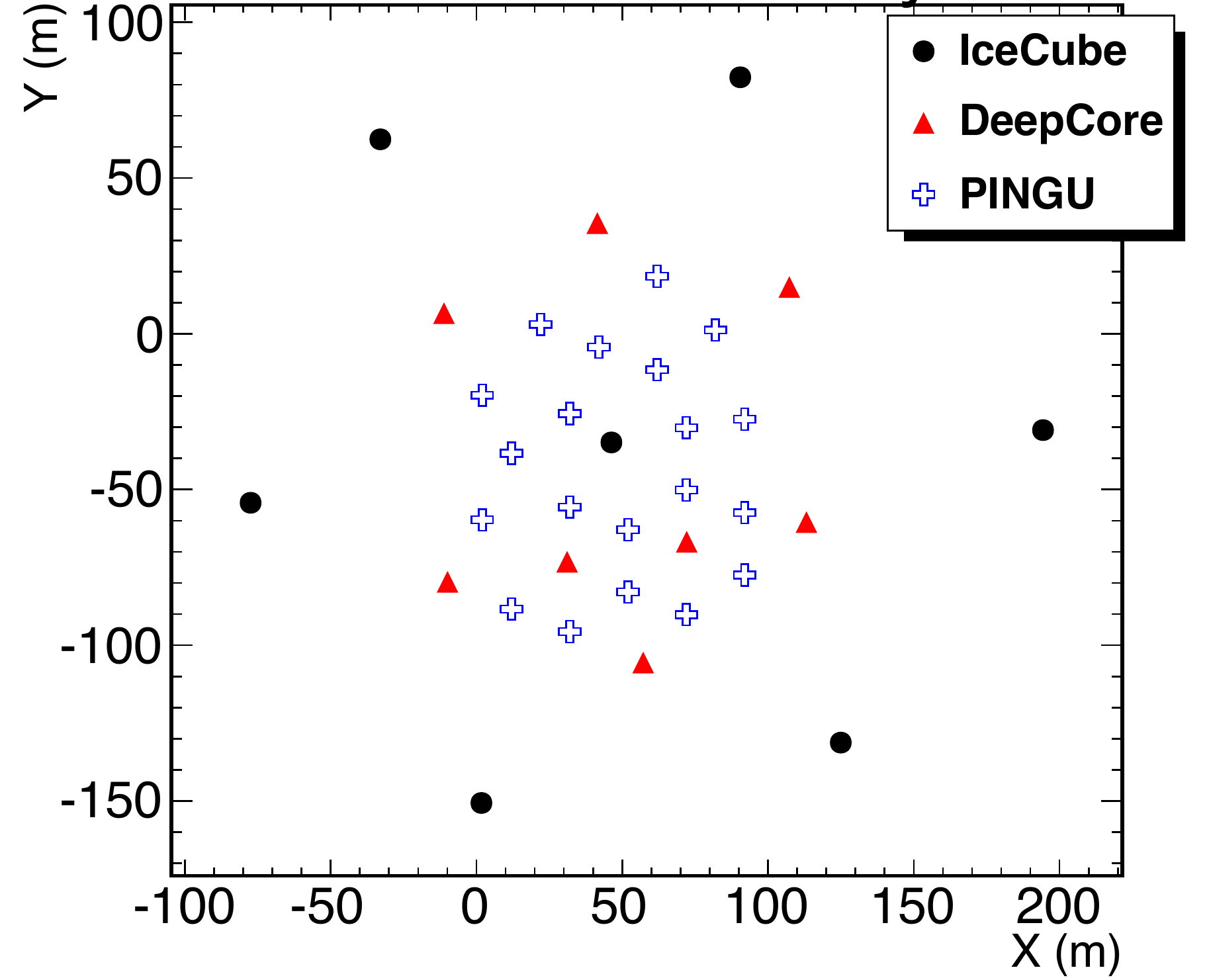}}
	\caption{ Possible geometry of a low energy infill detector PINGU with 20 
	additional strings instrumenting 5 Mtons of ice	in the center of IceCube. 
	An instrumented detector of that scale is being investigated with the goal to determine
	the mass hierarchy of neutrinos using upward going atmospheric neutrinos. }
	\label{fig:PINGU}
\end{center}
\end{figure}

Figure \ref{fig:PMTcoverage} illustrates the PMT coverage of 
existing and planned water Cherenkov detectors. 
The effective photon detection coverage (A$_{EPD}$) as defined earlier 
is divided by the instrumented or fiducial volume of the detectors 
and plotted versus the approximate energy threshold of the detector.  The PMT coverage 
scales quite well to the inverse of the threshold for a wide range of detectors. 
SuperK like detectors have an effective coverage of about 10$^4$m$^{2}$/Mton 
where high energy neutrino telescopes employ $\approx$0.1m$^{2}$/Mton.
IceCube's threshold is about $10^5$ times higher than Super-K's while
its effective photodetection coverage is about $10^{-5}$ times smaller. 
In the absence of absorption effects, such as the wall of a tank or the 
absorption in the medium, the inverse proportionality 
simply reflects that approximately the same number of photoelectrons 
is needed per event to perform physics analyses. 
IceCube's DeepCore has lowered the detector threshold to about 10\,GeV. 
Based on the progress with measurements of atmospheric neutrino oscillations,
IceCube and several interested groups not currently in the collaboration have begun  
investigating the design and capability of yet a 
more densely instrumented detector configuration. 
This "Precision IceCube Next Generation Underground" detector (PINGU) 
\cite{Grant_pingu_poster}
would consist of an instrumented volume of about 5 Mtons with an 
effective photodetection coverage per Mton of $\approx$50 times IceCube's
and $10^{-3}$ times that of Super-K. 
It would represent a serious move in closing the gap in Figure \ref{fig:PMTcoverage}.
This instrumentation density would be achieved by deploying an additional 20 strings 
of optical sensors instrumented with the same high quantum efficiency 
PMTs as IceCube's DeepCore detector.  A possible geometry of the additional strings is shown 
in figure \ref{fig:PINGU}.
Deployment of more complex R\&D modules such as the KM3NeT module
are being investigated by new groups.  

One of the primary science goals of PINGU is the determination of the neutrino mass hierarchy.
Detailed estimates about oscillation probabilities in the energy and zenith 
angle ranges of interest (3 - 20\,GeV, 120 - 180\,$^\circ$) have been 
presented by Akhmendov, Razzaque and Smirnov \cite{Akhmedov}, 
who also provide further references.  
They conclude that after 5 years of PINGU (20 string) operation in IceCube the significance of the determination of the hierarchy may range from $4\sigma$ to $11\sigma$ (without taking into account parameter degeneracies), depending on the accuracy of reconstruction of the neutrino energy and zenith angle.

For PINGU, the cost for instrumenting such a 5-Mton detector 
can be estimated based on IceCube's experience.  
It is worth noting that the ANTARES collaboration has published 
an analysis on atmospheric neutrino oscillations \cite{Antares_oscillations} 
and that the outlined strategy for neutrino oscillations
including mass hierarchy measurements may equally be pursued 
in water. 
While the signal situation is quite comparable between water and ice, it would be necessary to
investigate whether noise conditions and atmospheric muon rejection 
equally allow for measurements in the range of a few GeV using denser instrumentation.  
Given adequate instrumentation there is no obvious reason why that would not be possible.
Neutrino telescopes may contribute substantially to determining
parameters like the mass hierarchy.

\section{High-energy extensions: 100 PeV to 100 EeV}
\label{sec:high_energy}

At extremely high energies, in the range from 100\,PeV to 
100\,EeV, a cosmogenic neutrino flux is expected 
from the interaction of highest energy cosmic-ray protons 
in the cosmic microwave background. 
Predicted fluxes are in a range of approximately 1 event/year/km$^3$
or lower. 


\begin{figure}
\begin{center}
	\resizebox{\linewidth}{!}{\includegraphics{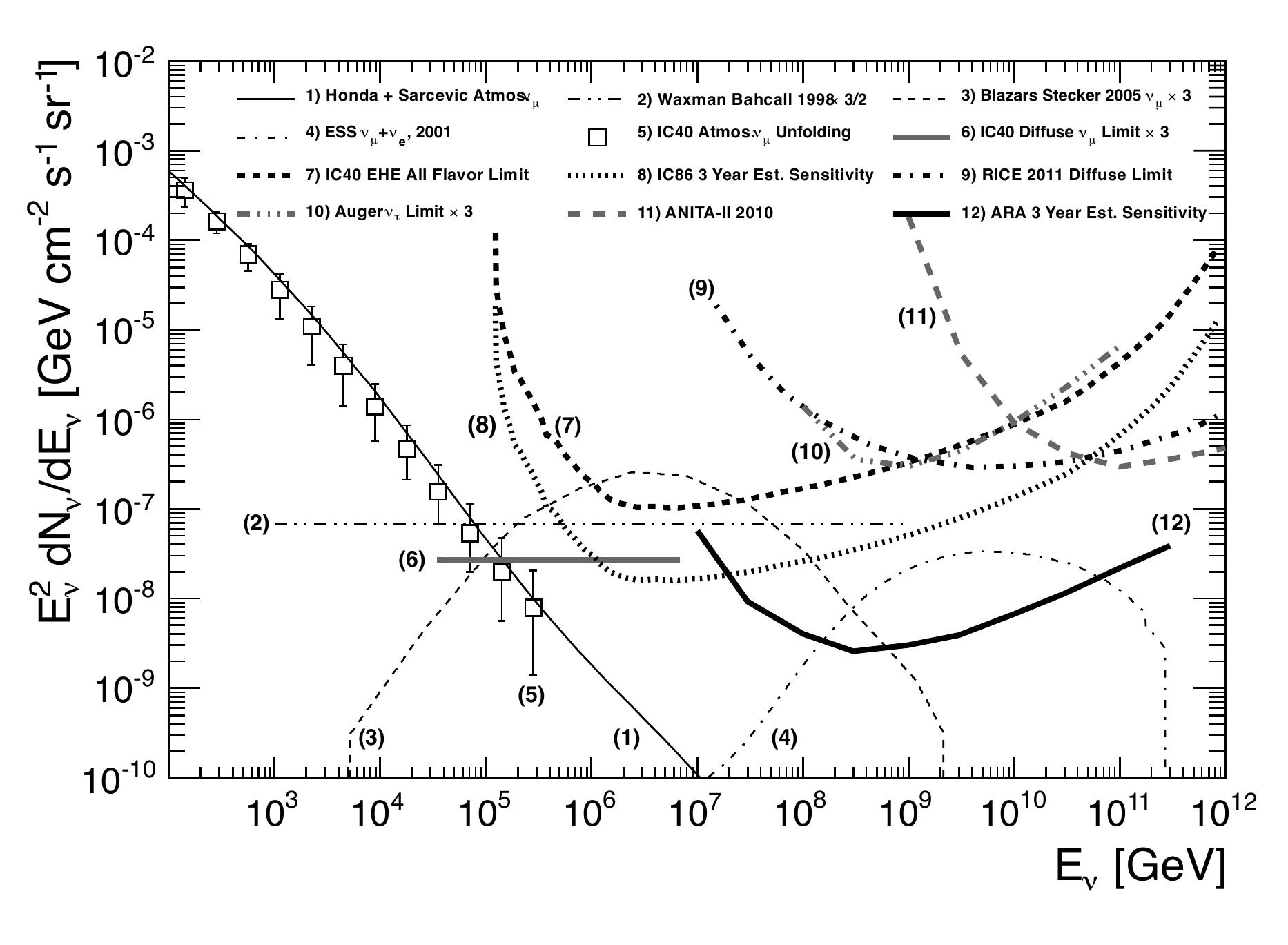}}
	\caption{ The measured atmospheric neutrino flux by IceCube is shown together with 
	several predictions of neutrino fluxes and upper limits by experiments: 
	1) Atmospheric neutrino flux by Honda 	\cite{Honda} + prompt by Sarcevic \cite{Sarcevic}, 
	2) Diffuse neutrino flux \cite{WaxmanBahcall}, 
	3) AGN Blazars \cite{Stecker}, 
	4) Cosmogenic neutrino flux \cite{ESS}, 
	5) IceCube atmospheric neutrino flux unfolded measurement \cite{IceCube-atmosph-IC40}, 
	6) IceCube 40, 1-year upper limit to diffuse neutrino flux \cite{IC40-diffuse}, 
	7) IceCube 40, 1-year upper limit to extremely high-energy neutrinos \cite{IceCube-EHE-IC40}, 
	8) IceCube 86, rough estimate of 3-year sensitivity (before this conference),
	9) RICE upper limit \cite{Rice}, 
	10) Auger 2-year limit x 3 \cite{Auger2012}, 
	11) ANITA upper limit \cite{Anita}, and 
	12) the Askaryan Radio Array (ARA) estimated 3-year sensitivity \cite{ARA}.  Differential limits are corrected for energy binning and flavor differences. }
	\label{fig:DiffuseFluxes}
\end{center}
\end{figure}

Cosmic rays have been measured to energies beyond 10$^{20}\,$\,eV. Ultrahigh-energy cosmic ray protons will interact with photon fields in the Universe; most prominently, every time a charged pion is produced, three neutrinos (muon neutrino, antineutrino, and an electron neutrino) are produced. This mechanism results in a cosmogenic neutrino predicted already in the 1960'ies by several authors.
Numerous, more detailed calculations have been presented since then. We show a reference model in Figure \ref{fig:DiffuseFluxes}, labeled as ESS \cite{ESS}. Uncertainties in the predictions include the cosmological evolution or distance distribution of the sources and the fraction of cosmic ray protons in the highest energy cosmic ray flux. The predicted fluxes are low. IceCube, which is optimized for lower energies, may have the best chance to see this flux with predictions around 1~event/year. An optical Cherenkov detector could be designed with greatly reduced 
photodetection coverage extending the scale in Figure \ref{fig:PMTcoverage}. 
However, in practice it would be not work very well due to requirements of 
installation and optical transmission.  
One could increase the spacing of strings to about 300 m, yet a very large-scale detector on the order of 1000~km$^{3}$sr acceptance would be too costly. In order to reliably detect this flux, other experimental strategies are needed that can be more feasibly optimized for this energy range.

\begin{figure}
\begin{center}
	\resizebox{0.95\linewidth}{!}{\includegraphics{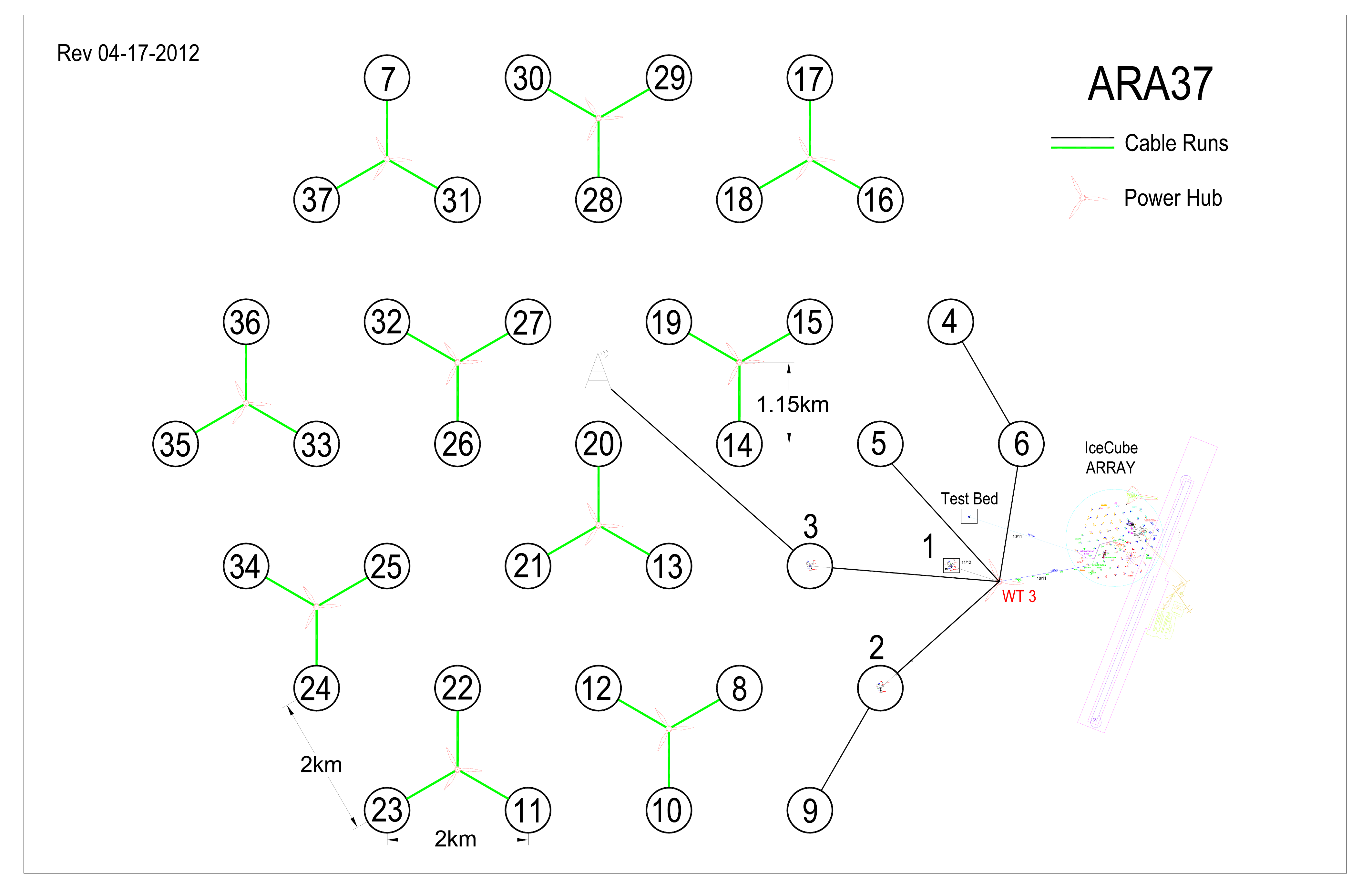}}
	\caption{Example of the proposed ARA radio neutrino detector geometry: 37 stations 
	would cover an area of about 200~km$^2$ of ice.  Hundreds of km$^3$ of target 
	volume are necessary to reach the desired sensitivities.
	See text for description of large radio detector configurations such as ARA and ARIANNA. 
	}
	\label{fig:ARAgeometry}
\end{center}
\end{figure}

An alternate detection mechanism was suggested as early as 1962 when G. Askaryan \cite{Askaryan} proposed that high-energy showers might produce coherent radio emission in dense media. These emissions would arise as an excess of negative charge builds up  as electrons are swept out along a relativistically advancing shower front (20\% more electrons than positrons when the shower is fully developed). The wavelength components of the broadband radiation from the motion of this net negative charge will add coherently for wavelengths that are large compared to the dimension of the charge distribution. The coherent emission is most pronounced in the frequency range from 1 GHz (on the Cherenkov cone) to 100 MHz (10$^\circ$ off the Cherenkov cone).
The large attenuation length of the cold glacial ice has been measured to be of order 1\,km in the relevant frequency range of 200 to 1000 MHz. This allows measuring UHE neutrino interactions in the deep ice with a few antennas located close to the surface of the 2.8 km thick ice sheet.

Several pioneering efforts have already been made to develop this approach, including RICE \cite{Rice}, ANITA \cite{Anita}, and early radio detection instrumentation in IceCube. 
Based on their experiences as well as the drilling and neutrino detector construction experience of IceCube at the South Pole, the Askaryan Radio Array (ARA) Collaboration has started to design, build and deploy prototypes of a detector array with the sensitivity to determine the cosmogenic neutrino flux. In its full configuration, illustrated in Figure \ref{fig:ARAgeometry} and described in ref. \cite{ARA}, the ARA-37 detector would consist of 37 detector stations covering an area of $200\,km^2$.  Each station consists of a cluster of 16 embedded antennas, deployed up to 200 m deep in four vertical boreholes placed with tens-of-meter horizontal spacing. Each such station is a fully functioning neutrino detector. 
For example, a neutrino interaction of $10^{18}$ eV can be detected up to distances of several km and in more than 2 km of depth, depending 
on the angle of the neutrino path relative to the detector.
As few as 16 antennas close to the surface would allow monitoring more then 10\,km$^3$ of ice for EeV neutrino interactions. 
ARA began its first deployments in the austral summer of 2011/12 and has continued deployment scheduled for the 2012/13 season. Based on first measurements, the \cite{ARA,ARA_Posters} the collaboration could
confirm the km-scale attenuation length for electromagnetic waves of the relevant frequency range in the massive cold ice sheet. It could also confirm that the electromagnetic environment is indeed very quiet and suitable for ARA detectors.

ARIANNA \cite{ARIANNA} is a project in an R\&D phase that plans to utilize the Ross Ice Shelf 
in Antarctica. The 600\,m thick shelf ice is relatively transparent to electromagnetic radiation at radio frequencies, and the water-ice boundary below the shelf creates a mirror to reflect radio signals from neutrino interactions in any downward direction. 
The baseline concept for ARIANNA consists of antenna stations arranged on a 100\,x\,100 square grid, separated by about 300\,m. Each station consists of a small group of cross-polarized antennas residing just beneath the snow surface and facing downwards. 
The station density is larger by more than a factor of 10 compared to ARA and the total surface area 
is envisioned to be 1000\,km$^2$.  
The shelf ice is warmer, and therefore less transparent, than the much colder ice at the 
South Pole; however, events are viewed from smaller distances on a denser grid 
such that the absorption should be adequate to be make full use of the ice sheet. 
A similar sensitivity is expected as for ARA-37.

As we can see from Figure \ref{fig:DiffuseFluxes}, which shows selected cosmogenic neutrino flux predictions and current experimental upper limits, the upper limits of experiments like ANITA, RICE, Auger and IceCube are approaching the predicted cosmogenic neutrino flux but are not close enough to test the model shown. 
The recent Auger result is based on Earth-skimming tau neutrinos with 2 years of lifetime. The IceCube result is based on one year of lifetime with only 50\% of the completed detector. In several years of lifetime, IceCube will have good chances of seeing events of the given reference flux. The most recent result from IceCube has been shown at this conference \cite{Ishihara_Nu2012}.
However, with sensitivity more than an order of magnitude greater than all existing experiments, ARA-37 or an experiment of similar sensitivity would be able to provide a definitive measurement of the cosmogenic neutrino flux. 
Such an experiment should observe between 10 and 100 events depending on the choice of model, 
most importantly the mass composition of cosmic rays at highest energies.

\section*{Acknowledgments}
I like to acknowledge the support from the U.S. National Science Foundation-Office of Polar Programs, the U.S. National Science Foundation-Physics Division and the University of Wisconsin Alumni Research Foundation. I like to thank D. Chirkin, J. Aguilar, Ch. Weaver, C. Kopper,  and  J. Koskinen and other IceCube collaborators for useful comments.  I thank U. Katz, M. de Jong, P. Sapienza for helpful comments on KM3Net;  Ch. Spiering and Z. Dzhilkibaev for useful comments on Baikal and 
S. Barwick for helpful comments on ARIANNA.



\end{document}